\documentclass[aps,prc,twocolumn,superscriptaddress,amsmath,showpacs,amssymb]{revtex4-1}
\usepackage{bm}
\usepackage[dvips,final]{graphicx}

\bmdefine{\bx}{x}
\bmdefine{\bn}{n}
\newcommand{\BM}{\begin{pmatrix}}
\newcommand{\EM}{\end{pmatrix}}

\renewcommand{\d}{\dagger}
\newcommand{\Lc}{\mathcal{L}}
\newcommand{\Mc}{\mathcal{M}}

\newcommand{\hphi}{\hat\varphi}
\newcommand{\hpsi}{\hat\psi}

\newcommand{\bra}[1]{\bigl\langle #1 \bigr|}
\newcommand{\ket}[1]{\bigl| #1 \bigr\rangle}
\newcommand{\braket}[2]{\bigl\langle #1 \big| #2 \bigr\rangle}

\newcommand{\intx}{\int\! d^3x\;}
\newcommand{\intxd}{\int\! d^3x'\;}
\newcommand{\intxxd}{\int\! d^3x\,d^3x'\;}

\newcommand{\ex}{\mathrm{ex}}

\begin{document}
\title {
Effective field theory of Bose--Einstein condensation of $\alpha$ clusters
and 
Nambu--Goldstone--Higgs states in $^{12}$C 
 }

\author{Y.~Nakamura}
\email{yusuke.n@asagi.waseda.jp}
\affiliation{Department of Electronic and Physical Systems, 
Waseda University, Tokyo 169-8555, Japan} 
\affiliation{Nagano Prefectural Kiso Seiho High School, Nagano 397-8571, Japan} 
\author{J.~Takahashi}
\email{j.takahashi@aoni.waseda.jp}
\affiliation{Department of Electronic and Physical Systems, 
Waseda University, Tokyo 169-8555, Japan} 
\author{Y.~Yamanaka}
\email{yamanaka@waseda.jp}
\affiliation{Department of Electronic and Physical Systems, 
Waseda University, Tokyo 169-8555, Japan} 

\author{S.~Ohkubo}
\email{shigeo.ohkubo@rcnp.osaka-u.ac.jp}
\affiliation{Research Center for Nuclear Physics,
 Osaka University,  Ibaraki,
Osaka 567-0047, Japan}
\affiliation{University of Kochi,  Kochi 780-8515, Japan }

\date{\today}

\begin{abstract}
\par
An effective field theory of $\alpha$ cluster condensation is formulated as 
a spontaneously broken symmetry in quantum field theory to understand the 
{\it raison d'\^etre} and nature of the Hoyle and $\alpha$ cluster states
 in $^{12}$C. The Nambu--Goldstone and Higgs mode operators 
in infinite systems are replaced with a pair of canonical operators 
whose Hamiltonian gives rise to discrete energy states in addition to 
the Bogoliubov--de Gennes excited states. The calculations reproduce well 
the experimental spectrum of the  $\alpha$ cluster states. 
The existence of the Nambu--Goldstone--Higgs states is demonstrated and crucial.
The $\gamma$ decay transitions are also obtained.
  \end{abstract}

\maketitle
\par

\section{Introduction}
%% main text
Alpha cluster condensation in nuclei has attracted much attention since 
the observation of Bose--Einstein condensation (BEC) of trapped cold 
atoms \cite{Cornel2002}.
%
% More than fifty years ago Morinaga \cite{Morinaga1956} proposed that
% the Hoyle state is a band head state with a linear chain  structure 
%of three $\alpha$  clusters.
In  $^{12}$C,  the three-$\alpha$ structure was most thoroughly investigated by
 Uegaki {\it et al.} \cite{Uegaki1977}, who
  showed that the $0_2^+$ state at  an excitation energy $E_x$ of 7.654 MeV,
  the Hoyle state, which is crucial for  nucleosynthesis, the evolution
   of stars, and the emergence of life, has a dilute structure in a new  
 ``$\alpha$-boson gas phase''   and clarified  the systematic existence of 
  a ``new phase'' of three $\alpha$ clusters above the $\alpha$ threshold.  
The  Hoyle state has been  extensively  studied theoretically 
\cite{Tohsaki2001,Yamada2004,Ohkubo2004,Yamada2005,Kurokawa2005,Kurokawa2007,
Kanada2007,Chernykh2007,Roth2011,Epelbaum2012,Dreyfuss2013}
and experimentally 
\cite{Freer2009,Itoh2011,Zimmerman2011,Zimmerman2013,Itoh2013,Freer2014,Freer2011,Ogloblin2014},
 and has been considered widely as an  $\alpha$ cluster condensate.
It has a
gas-like structure  with  a dilute matter distribution of three-$\alpha$
clusters, 70\% of which are 
 in the $0$s state \cite{Yamada2005}.
However, no firm evidence of BEC, such
as superfluidity,   has been found.

\par
In $^{12}$C, all the excited states except the $2_1^+$ state at 4.44 MeV appear
 above the $\alpha$ particle threshold (7.367 MeV).
 Recently,   $\alpha$
 cluster states above the Hoyle state, which are also  candidates for  an $\alpha$ cluster
 condensate,
 that is,  the $0_3^+$ state at 9.04 MeV, $0_4^+$ state at 10.56 MeV,
 $2_2^+$ state  ($\sim$9.75 MeV) \cite{Itoh2011,Itoh2013,Freer2009,Zimmerman2011,Zimmerman2013},
 and  $4_1^+$ state ($\sim$13.3 MeV)  \cite{Freer2011,Ogloblin2014} 
have been observed.
To date, studies using $\alpha$ cluster models
 \cite{Yamada2005,Kurokawa2005,Kurokawa2007}
 and  {\it ab initio}
calculations  \cite{Kanada2007,Chernykh2007,Roth2011,Epelbaum2012,Dreyfuss2013}
 explain the Hoyle state and the excited gas-like 
 states as collective states of $\alpha$ clusters or nucleons in
 {\it configuration space}.
Collective motions arise owing mostly to spontaneously broken symmetries (SBSs)
in the configuration space, such as rotational and translational ones,
or in the gauge space \cite{Ring1980, BalaizotRipka}.
The  BEC of $\alpha$ clusters  is a manifestation
 of the SBS of the global phase. It would be
 difficult from the standpoint of
traditional $\alpha$ cluster models or
{\it ab initio} calculations  to conclude that
 BEC is truly  realized, because it is not clear then what type of
{\it symmetry} is broken for the Hoyle state and the $\alpha$ condensate
 states above it.

In the study of $\alpha$ cluster condensation,
it is important to treat the SBS of the global 
phase on the basis of quantum field theory because of its
unifying view and underlying principle. SBS is 
ubiquitous \cite{Watanabe2012}; when it occurs,
a Nambu--Goldstone (NG) mode (phason) appears 
according to the NG theorem \cite{Nambu1961,Goldstone1961}, and 
a Higgs (amplitude) mode
(amplitudon) usually accompanies it.  For example, in infinite superconducting systems,
 the NG mode \cite{Kadowaki1998}, which  is eaten by the plasmon, and the Higgs mode
 \cite{Littlewood1981,Varma2001} have been observed.
 For systems with a finite particle number, both the NG and Higgs modes have
 been confirmed   in superfluid nuclei
as  a  pairing rotation and pairing vibration, respectively \cite{Broglia1973}.
The observation of the Higgs boson  in particle physics \cite{ATLAS2012}
has stimulated a search for Higgs modes in other phenomena, including a recent 
experiment on Higgs mode
excitation   in a superconductor using a terahertz pulse \cite{Matsunaga2013}.
It is intriguing to reveal the emergence of the NG and Higgs
modes theoretically in an $\alpha$ cluster condensate 
and to observe them experimentally. Because the system is finite in size and particle
number, they would manifest themselves not as particle excitations but as
   resonant  states with discrete energy levels.
From this viewpoint, Ref.\cite{Ohkubo2013} discussed a possible
  emergence of such states
for an $\alpha$ cluster condensate in $^{12}$C and $^{16}$O 
qualitatively.

% purpose of the paper
 The purpose of this paper is to show for the first time that the
dilute  excited $\alpha$ cluster states, the Hoyle state and those above it,
can be understood as new discrete states that follow naturally
in the formulation of quantum field theory \cite{NTY}, 
called the interacting zero mode formulation (IZMF in short), 
for BEC of $\alpha$ clusters
in terms of the field equation, canonical commutation relations (CCRs),
 and global gauge invariance. 

This paper is organized as follows. In Sect.~\ref{sec-IZMF}, 
the IZMF for BEC of trapped cold atoms is extended to BEC of $\alpha$ clusters.
In Sect.~\ref{sec-Parameters}, we introduce a phenomenological model 
of $\alpha$ clusters, in which $\alpha$ particles are trapped by a harmonic potential
and the $\alpha$-$\alpha$ interaction is described by a phenomenological 
Ali--Bodmer potential \cite{Ali1966}.  Then, the strengths of the harmonic potential and
the repulsive potential in the Ali--Bodmer potential are the key parameters in our analysis.
We calculate the energy levels, adjusting the two parameters, and compare them with
the observed $\alpha$ cluster states. The $\gamma$ decay transition probabilities 
are calculated in Sect.~\ref{sec-gamma}. Sect.~\ref{sec-Summary} is devoted to the summary.

\section{Formulation of quantum field theory of Bose-Einstein condensation for $\alpha$ clusters}
\label{sec-IZMF}

 First, we clarify from quantum field theory for the  $\alpha$ cluster condensate
that  the canonical operators \cite{NTY}, which replace
the NG and Higgs mode  operators in  infinite systems with SBS, 
emerge and that the spectrum of their quantum mechanical system is discrete.
%%%%%%%%%%%%%%%%%%%%%%%%%%%%%%%%%%%%%%%%%%%%%%%%%%%%%%%%%%%%%%%%%%

We start with the following Hamiltonian for the $\alpha$ cluster system
described by the field operator ${\hat \psi}$:
\begin{align}\label{Hamiltonian}
	&\hat{H}=\intx \hpsi^\d(x) \left(-\frac{\nabla^2}{2m}+
	V_\ex(\bx)- \mu \right) \hpsi(x) \notag\\
		&\,\,+\frac12 \intxxd \hpsi^\d(x) 
		\hpsi^\d(x') U(|\bx-\bx'|) \hpsi(x') \hpsi(x) \,,
\end{align}
where $m$ and $\mu$ denote the mass of the $\alpha$ particle 
and the chemical potential, respectively. The external isotropic 
confinement potential $V_\ex(\bx)$ is introduced in a phenomenological manner
that will be discussed later.  
The interaction potential $U(r)$ is the sum of the nuclear $\alpha$--$\alpha$
 potential,  $V_{\alpha-\alpha}^{\rm Nucl}(r)$, 
and the Coulomb potential, $V_{\alpha-\alpha}^{\rm Coul}(r)$. 
We set $\hbar=c=1$
throughout this paper.

Assuming $\alpha$ condensation, namely, the broken phase,
 we divide  $\hpsi$ into a condensate c-number 
component $\xi$ and an excitation component $\hphi$ using the criterion $\bra0 \hpsi \ket0=\xi$. 
The order parameter $\xi$ is taken to be stationary, isotropic, and real, and is 
normalized to the condensed particle number as $\intx \xi^2(\bx) = N_0$, 
where we fix $N_0=3$ for ${}^{12}\mathrm{C}$ below.
 The Hamiltonian (\ref{Hamiltonian}) is rewritten in 
terms of $\hphi$ as $\hat{H} = \hat{H}_2 + \hat{H}_{3,4}$, where
\begin{align}
\label{H2}
&\hat H_{2} = \frac12 \intxxd \BM\hphi^\d(x)  & -\hphi(x) \EM\notag\\
&\qquad\qquad \times\BM \Lc(\bx,\bx') & \Mc(\bx,\bx') \\
-\Mc(\bx, \bx') & -\Lc(\bx,\bx') \EM
\BM\hphi(x')  \\\hphi^\d(x') \EM \,,\\
&\hat H_{3,4} =\frac 12 \intxxd U(|\bx-\bx'|) \notag\\
&\quad \times \left[\left\{
2 \xi(\bx')+\hphi^\d(x')\right\}\hphi^\d(x)\hphi(x)\hphi(x') +{\rm h.c.} \right]
\,,
\end{align}
with
\begin{align}
V_H(\bx) &= \intxd U(|\bx-\bx'|)\xi^2(\bx')\,, \\
\Mc(\bx,\bx') &= U(|\bx-\bx'|)\xi(\bx) \xi(\bx')\,,\\
\Lc(\bx,\bx') &= \delta(\bx-\bx') \bigl(-{\nabla^2}/{2m}+V_\ex(\bx) \notag\\
&\hspace{1.5cm}-\mu + V_H(\bx) \bigl) +\Mc(\bx,\bx')\,.
\end{align}
The requirement that the $\hphi$-linear term in ${\hat H}$ must vanish leads 
to the Gross--Pitaevskii equation \cite{GP}
\begin{equation}\label{GP}
\left( -{\nabla^2}/{2m}+V_\ex(\bx) -\mu + V_H(\bx) \right) \xi(\bx) = 0 \,.
\end{equation}
According to the method developed in cold atomic physics, $\hphi$ is expanded as 
\cite{Matsumoto2, Lewenstein}
\begin{equation} \label{fieldexpansion}
\hphi(x) = \hphi_\ex(x)  -i \hat Q(t) \xi(\bx) + \hat P(t) \eta(\bx)\,.
\end{equation}
The field $\hphi_\ex(x)$ is expanded as 
$\hphi_\ex(x) = \sum_{\bn}\left[ \hat a_{\bn}(t) u_{\bm n}(\bx) 
+\hat a_{\bm n}^\d(t) v_{\bn}^{*}(\bx) \right] $\,, 
where $u_\bn$ and $v_\bn$ are the elements of the Bogoliubov--de Gennes (BdG) 
eigenfunction
\cite{Bogoliubov, deGennes}, 
\begin{equation} \label{eq:BdG}
\intxd \BM \Lc& \Mc \\-\Mc & -\Lc\EM
\BM u_\bn \\ v_\bn\EM  = \omega_\bn \BM u_\bn \\ v_\bn\EM\,,
\end{equation}
with a normalization condition $\intx \bigl[ |u_\bn|^2 -|v_\bn|^2\bigr] = 1\,$.
The isotropic $\xi$ implies $\bn = (n,\, \ell,\, m)$, 
a triad of the main, azimuthal, and magnetic quantum numbers. 
In Eq.~(\ref{fieldexpansion}), $\xi$ is the element 
of the BdG eigenfunction belonging to zero eigenvalue,
and $\eta$ is its adjoint function, calculated as 
\begin{align}\label{eq:eta}
\eta(\bx)= \frac{\partial} {\partial N_0} \xi(\bx)\,,
\end{align}
with a normalization condition $\intx \bigl[\xi^\ast \eta+ \eta^\ast \xi\bigr] = 1\,$.
The CCR of $\hpsi$ and $\hpsi^\dagger$ yields
$
[\hat a_{\bm n} ,\hat a_{\bm n'}^\d] = \delta_{\bm n \bm n'} \,,\;
[\hat Q ,\hat P] = i \,,\;  
\text{(otherwise)} = 0 \,.$
The pair of canonical operators ${\hat Q}$ and ${\hat P}$, 
which are associated with the eigenfunctions
with zero eigenvalue and stem from the SBS of the global phase,
are counterparts of the NG and Higgs mode operators in general infinite systems.
 The use of the mode operators in our finite 
 system does not diagonalize the unperturbed Hamiltonian and also causes 
singular behavior, whereas that of ${\hat Q}$ and ${\hat P}$ is free from these
 difficulties. 
We call $({\hat Q}\,,\,{\hat P})$ and 
the subspace of states operated by them the Nambu--Goldstone--Higgs (NGH)
 operators and NGH subspace 
(or simply zero mode operators and zero mode subspace), respectively.
 The excitation mode created by
${\hat a}_\bn^\dagger$ is referred to as the BdG mode.
We note that the NGH operators exist in our finite model of superfluid type 
irrespective of the fact that the Higgs mode is absent 
in non-relativistic infinite models of this type \cite{Varma2001}.

Let us seek the vacuum $\ket0$, 
%or ground state, 
with which we identify
 the Hoyle state. A naive choice of the unperturbed Hamiltonian 
would be ${\hat H}_2$,  because the system is a dilute, weakly interacting gas-like
one, so the higher powers of $\hphi$, $\hat H_{3,4}$ could be ignored in the
leading order.
Substituting Eq.~(\ref{fieldexpansion}) into Eq.~(\ref{H2}), we obtain
$\hat{H}_2 = {I\hat{P}^2}/{2} + \sum_\bn \omega_\bn \hat a_\bn^\d \hat a_\bn \,,$
with $I = \partial \mu / \partial N_0$. The Hamiltonian of the NGH operators, which has
 the free-particle  form and therefore a continuous spectrum, causes serious defects,
that is, the non-existence of a stationary normalized vacuum and 
the diffusing phase of $\xi$ \cite{Lewenstein}.

In the traditional formulations such as in Refs.~\cite{Lewenstein} and \cite{Marshlek1969}, the linear expansion is replaced with the approximate non-linear expansion,
\begin{equation}
{\hat \psi}(x) \simeq  
e^{-i \hat Q(t) }\left\{\xi(\bx) + \hat P(t) \eta(\bx)+ \hphi_\ex(x)\right\}\,   
\label{NonLinExp}
\end{equation}
under the assumption of small ${\hat Q}$.  The authors of Ref.~\cite{Marshlek1969} specified the global properties of ${\hat Q}$ and ${\hat P}$, identifying  them 
as the azimuth angle and angular momentum operators, respectively,
on the ground of the expression~(\ref{NonLinExp}) although its validity is restricted to
small ${\hat Q}$\,. As a result, the spectrum of ${\hat P}$ and consequently 
that of the Hamiltonian become discrete, so one does not encounter
the defects in the preceding paragraph. However, we can not accept 
the non-linear expansion~(\ref{NonLinExp}) from the standpoint of quantum field theory
because it
violates the CCR of ${\hat \psi}$ and ${\hat \psi^\dagger}$\,. As will be given just below, 
we insist on the linear expansion~(\ref{fieldexpansion}), that is, 
the CCR, but introduce the non-linear 
unperturbed Hamiltonian instead of  the bilinear
one.

To avoid the defects mentioned above, 
 a modified unperturbed Hamiltonian \cite{NTY},
which retains the nonlinear terms of $\hat Q$ and $\hat P$ in $\hat H_{3,4}$\,,
 has been proposed, because
it is unfounded to neglect them, unlike the higher powers 
of the BdG modes. Concretely, we replace the term ${I\hat{P}^2}/{2}$ above 
with
\begin{align} \label{eq:HuQP}
&\hat H_u^{QP} = - \left(\delta\mu + 2C_{2002} + 2C_{1111} \right) \hat P+
 \frac{I-4C_{1102}}{2}\hat P^2 \notag \\
&\,\, + 2C_{2011}\hat Q\hat P\hat Q 
+ 2C_{1102}\hat P^3 + \frac{1}{2}C_{2020}\hat Q^4 -2C_{2011}
\hat Q^2 \notag \\
&\,\,+ C_{2002}\hat Q\hat P^2\hat Q + \frac{1}{2} C_{0202}\hat P^4 \,,
\end{align}
where $C_{iji'j'}= \int\! d^3x d^3x' U(r)  \xi^{i}(\bx)\eta^{j}(\bx)
\xi^{i'}(\bx')\eta^{j'}(\bx')$ with $r=|\bx-\bx'|$ ,
and $\delta \mu $ is to be determined self-consistently 
to satisfy the criterion $\bra0\hpsi\ket0=\xi\,$.
The fact that the spectrum of $\hat H_u^{QP}$ is discrete is especially significant.
It is implicitly postulated in the introduction of Eq.~(\ref{eq:HuQP})
that the unperturbed state of the total system is factorized 
as $\ket{\Psi}\ket{\cdot}_\ex  $, 
where $\ket{\Psi}$ and $\ket{\cdot}_\ex$ are a wave function in the NGH subspace
and a Fock state associated with $\hat a_\bn$, respectively. 
Accordingly, all the cross terms such as $ {\hat a}_{{\bn}} 
{\hat Q}{\hat P}$ are included in the interaction Hamiltonian and should be 
treated perturbatively. The unperturbed vacuum $\ket{0}\,$, 
which is identified with the Hoyle state,
is now given by $ \ket{\Psi_0}\ket{0}_\ex $\,, where $\ket{\Psi_0}$
is the ground state in the NGH (zero mode) eigenequation 
\begin{align} \label{eq:HuQPeigen}
\hat H_u^{QP} \ket{\Psi_\nu} = E_\nu \ket{\Psi_\nu}\qquad 
(\nu=0,1,\cdots)\,.
\end{align}
The excitation in the NGH subspace
is a new and original concept and our prediction, for which  the adoption of 
the non-quadratic Hamiltonian in Eq.~(\ref{eq:HuQP}) is crucial  \cite{NTY}. 
Note that this 
excitation does not change  the value of the angular momentum $J$
because the NGH operators carry no quantum number in configuration space.
The states $\ket{\Psi_\nu}\ket{0}_\ex
\,\, (\nu=1,2,\cdots)$, which have gap energies from the Hoyle state 
$E_\nu - E_0 \;$, are referred to as the {\it NGH states\/} below.
 The BdG excitation energy $\omega_\bn$ 
is measured from the energy of the Hoyle state, 
and the state $\ket{\Psi_0}({\hat a_\bn^\dagger}\ket{0}_\ex)$ is termed
the {\it BdG state\/}. Its experimental  $J$
is given by the azimuthal quantum number $\ell$ of $\bn$. 
Solving the coupled system of the GP eq.~(\ref{GP}), BdG eq.~(\ref{eq:BdG}) with 
Eq.~(\ref{eq:eta}), and NGH (zero mode) eq.~(\ref{eq:HuQPeigen}),
we obtain theoretical predictions that can be compared with experimental data, as shown below.

\begin{figure}[t]
	\includegraphics  [width=8.0cm] {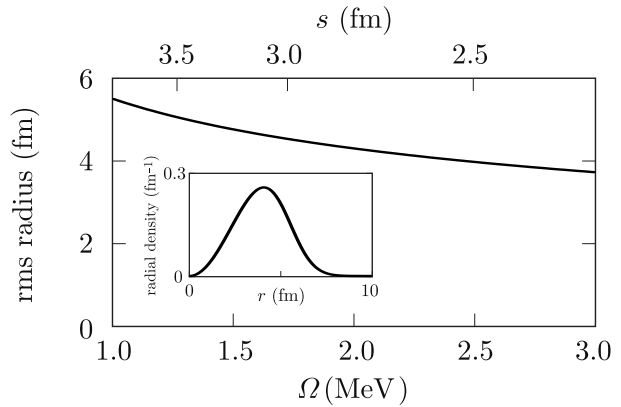}
	\caption{ Calculated ${\bar r}$
	 as a function of $\Omega$ 
 and  the radial density distribution for $\Omega$=2.14 MeV (inset) of the Hoyle state.
 Upper horizontal axis indicates rms radius $s=\sqrt{3/2 m\Omega}$ of the $0$s 
orbit of the external harmonic oscillator
potential.
	}
	\label{fig:rms_density}
\end{figure}

\section{Parameters and numerical calculations}\label{sec-Parameters}

In the calculations, we take a phenomenological Ali--Bodmer potential 
for 
$V_{\alpha-\alpha}^{\mathrm{Nucl}}(r)$, which is
characterized by the four parameters \cite{Ali1966},
\begin{align}
V_{\alpha-\alpha}^{\mathrm{Nucl}}(r)=V_r\, e^{-\mu_r^2 r^2}- V_a\, e^{-\mu_a^2 r^2}
\,,
\end{align}
where $V_r$ and $V_a$ are the strengths of the repulsive
and attractive parts, respectively, and $\mu_r$ and $\mu_a$ the corresponding inverse ranges.
This potential was obtained by fitting the s-wave phase
shifts of $\alpha$--$\alpha$ scattering and has been used in three-$\alpha$ cluster 
structure studies of $^{12}$C \cite{Yamada2004}. 
It has been a well-known fact that the Ali--Bodmer local potential does not reproduce the binding energy of the ground state and the Hoyle state.  The attraction of the Ali--Bodmer potential is too weak for the three alpha system. One way to reproduce the correct binding of these states is to introduce a strong three-body attracting force \cite{Yamada2004,Ogasawara1976,Lazauskas2011}.
Alternatively, in this paper, we introduce an external harmonic potential $V_{\rm ex}(r)=m\Omega^2 r^2/2$ which mimics the three-body attracting force to bind the Hoyle state. Here, $\Omega$ is a fit parameter which corresponds to the strength of the three-body force. The introduction of the external potential makes the theoretical analysis simpler without losing the self-binding essence. If we took only the interaction among the alpha particles, the original translational symmetry would be spontaneously broken in the formation of the nucleus. Then we would have an additional NG mode associated with the translational symmetry in addition to the one of the phase symmetry. To avoid this complexity, we explicitly break the translational symmetry by introducing the external potential.
The Coulomb potential, $V_{\alpha-\alpha}^{\mathrm{Coul}}(r)$, is taken as $(4e^2/r)\mathrm{erf}(\sqrt{3}r/2b)$, where the size parameter of the $\alpha$ particle $b$ is 1.44 fm. 

We attempt to calculate the rms radius, denoted by ${\bar r}=\sqrt{\langle r^2 \rangle}$, 
and density profile of 
the Hoyle state from $\xi(\bx)$\, taking the parameter set $d_0$ in Ref.~\cite{Ali1966}
with the proviso that the parameter $V_r$ decreases slightly from 500 MeV to 422 MeV,  
which is consistent with the finding in Ref.\cite{Tohsaki1980} that the $\alpha$--$\alpha$ interaction in the
three-$\alpha$ system is more attractive than that determined in free
$\alpha$--$\alpha$ scattering. The results are shown in Fig.~\ref{fig:rms_density}.
The Hoyle state is found to be dilute for all the $\Omega$ values.
The peak position of the radial density distribution, located around 4 $\sim$ 5 fm,
and ${\bar r}$ are not very sensitive to $\Omega$.

The coexistence of concentration by the trapping potential and 
repulsion by the self-interaction is crucial for a stable BEC of trapped cold atoms.
As a typical counterexample, the trapped BEC of attractively interacting atoms collapses.
We therefore regard $\Omega$ and $V_r$ as the key parameters in our study, and fit 
them below with the fixed $V_a=130$\,MeV, $\mu_a=0.475$\,fm$^{-1}$, and $\mu_r=0.7$\,fm$^{-1}$ 
in the parameter set $d_0$~\cite{Ali1966}.

\section{Energy spectrum}\label{sec-Energy}

%% Fig.2
\begin{figure}[t]
	\includegraphics [height=6.5cm]{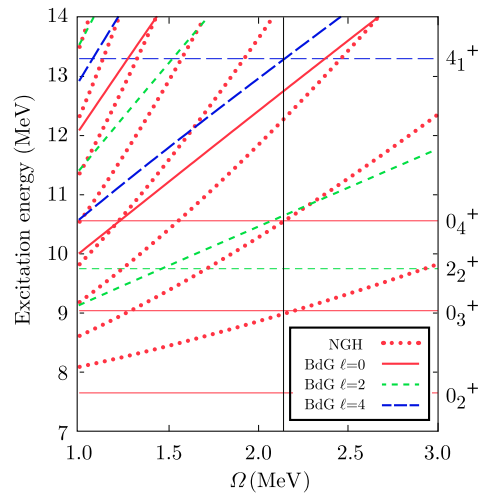}
	\caption{(Color online) NGH (dotted lines) and BdG (solid or dashed lines) 
		 excitation energies with $\ell=0$ 
		(red), 2 (green), and 4 (blue) as
		a function of $\Omega$ when $V_r=422$\,MeV is fixed. 
		The horizontal lines indicate the excitation energies of the observed
 $\alpha$ cluster states  in $^{12}$C
 \cite{Itoh2011,Freer2009,Zimmerman2011,Zimmerman2013,Itoh2013,Freer2011}, 
		and the vertical line is a guide to the  eye.
	}
	\label{fig:Omega_ex_E}
\end{figure}
\begin{figure}[tb]
	\includegraphics[height=4.5cm]{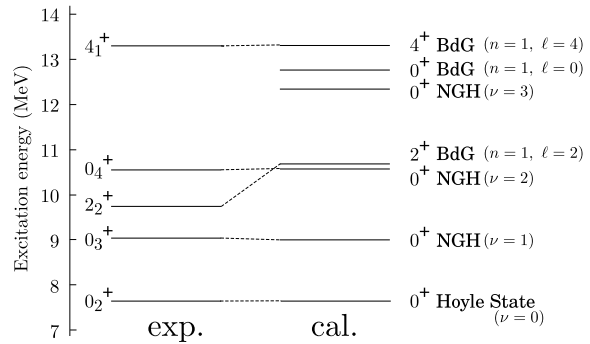}
	\caption{The calculated energy levels for 
		parameter set A ($\Omega=$2.14 MeV, $V_r=$422 MeV),
        compared with the observed $\alpha$ cluster states in $^{12}$C
 \cite{Itoh2011,Freer2009,Zimmerman2011,Zimmerman2013,Itoh2013,Freer2011}.
 	}
	\label{fig:12Cenergy}
\end{figure}

First we use only the existing experimental energy levels to determine 
the two fit parameters $\Omega$ and $V_r$\,.The $\Omega$-dependence
of the calculated energy levels is given in Fig.~\ref{fig:Omega_ex_E}. 
Fig.~\ref{fig:12Cenergy} shows 
the calculated  energy levels for the best fitting parameters 
$\Omega=2.14$\,MeV and $V_r=422$\,MeV, which is referred to as the parameter set A,
in comparison with the observed $\alpha$ cluster states. The calculated ${\bar r}$ of 0$_2^+$
is 4.21fm, which is comparable with the calculations in Refs.~ \cite{Uegaki1977,Tohsaki2001,Yamada2005,Kurokawa2007}.
The agreement between  the calculated  and experimental
energy levels is  good, and the  order of the  levels is correctly reproduced.
Our calculation reproduces the two  $0^+$ NGH states $(\nu=1,2)$\,, 
which correspond well to the $0_3^+$ at 9.04 MeV and $0_4^+$ at 10.56 MeV, respectively.
The existence of the NGH states is critical for the assignments,
because there is no BdG state with $\ell=0$ near the energies
of $0_3^+$ and $0_4^+$. Then, quite naturally, the excitations $2_2^+$ and $4_1^+$ are 
identified as the BdG states with $\ell=2, 4$.
All the observed positive parity states are well reproduced as  BEC states of $\alpha$ clusters.
This shows that the present field theory is useful even for a few-body  system.

In  Figs.~\ref{fig:Omega_ex_E} and \ref{fig:12Cenergy}, 
the calculation shows two $0^+$ states around $12.5 \mathrm{MeV}$, 
where no corresponding excitation has been established experimentally yet. 
These are the NGH state with $\nu=3$, $\ket{\Psi_3}\ket{0}_\ex$, and the BdG state $\ket{\Psi_0}
({\hat a}^\dagger_{100} \ket{0}_\ex)$, denoted simply as $\ket{\rm h}$ and $\ket{\rm BdG}$,
respectively. Because the energy difference between the two states is small and the 
interaction Hamiltonian allows mutual transitions, 
 they mix with each other to make new two energy eigenstates.
A rough estimation of diagonalizing ${\hat H}$  
%in which the interaction potential $U$ is approximated as
%a delta-function type and the total Hamiltonian is diagonalized
in the subspace of $\ket{\rm h}$ and $\ket{\rm BdG}$ gives  the eigenstates, 
$0.98\ket{\rm BdG}-0.18 \ket{\rm h}$
 with an energy of $12.60 \mathrm{MeV}$ and $0.98\ket{\rm h}+0.18 \ket{\rm BdG}$ with 
$12.20 \mathrm{MeV}$\,. The mixing is not remarkable here,
but is generally sensitive to the energy difference.
We also note that doubly excited states, e.g., 
$\ket{\Psi_1} ({\hat a}^\dagger_{02m} \ket{0}_\ex)$ are possible.

\begin{figure}[t]
	\includegraphics [width=7cm]{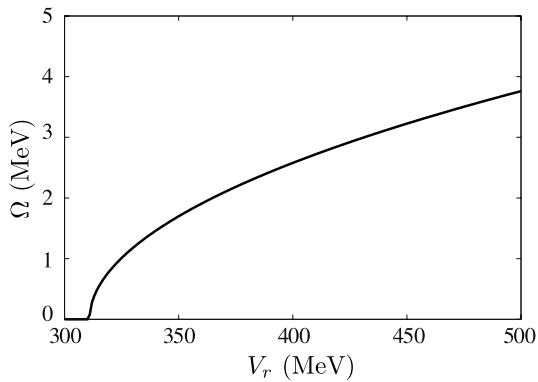}
	\caption{$V_r$-$\Omega$ plot for ${\bar r}=3.8\, \mathrm{fm}$}
	\label{fig:vr_Omega_fixed_rms_38}
\end{figure}

\begin{figure}[t]
	\includegraphics [height=6.5cm]{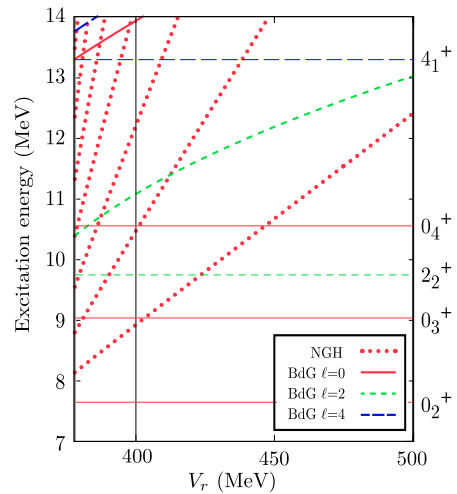}
	\caption{(Color online) NGH (dotted lines) and BdG (solid or dashed lines) 
		excitation energies with $\ell=0$ 
		(red), 2 (green), and 4 (blue) as
		a function of $V_r$ for fixed ${\bar r}=3.8\, \mathrm{fm}$. 
	}
	\label{fig:ex_E_fixed_rms_38}
\end{figure}

\begin{figure}[tb]
	\includegraphics [height=4.5cm]{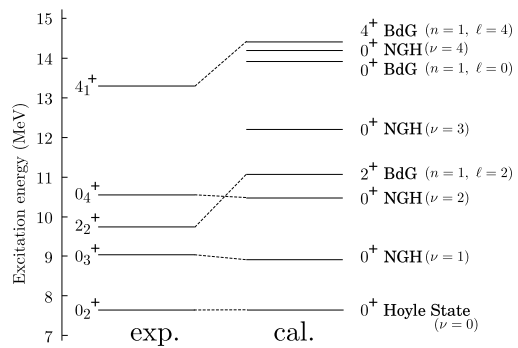}
	\caption{The calculated energy levels for parameter set B ($\Omega=$2.58 MeV, $V_r=$400 MeV).
 	}
	\label{fig:energyB}
\end{figure}

Next, we try to determine the parameters $\Omega$ and $V_r$ in another way. 
First of all, because the observed energy levels have large widths,
 their fitting is not very useful. We add the rms radius of the Hoyle 
state, depending on the wave function, as an object to be fitted. 
The value of ${\bar r}= 4.21\, \mathrm{fm}$, calculated for the parameter set A,
is rather large, compared with the typical range $3.3\, \sim \, 3.8\, \mathrm{fm}$ obtained
in other $\alpha$ cluster model calculations \cite{Tohsaki2001,Fukushima1977,Uegaki1977,Kanada2007,Matsmura2004}, and the  values around $2.9\, \mathrm{fm}$ 
 estimated from inelastic scattering from the Hoyle state \cite{Danilov2009}. 
We seek values of the parameters
that give energy levels consistent with the observed energy levels, fixing 
${\bar r}$. The plot in Fig.~\ref{fig:vr_Omega_fixed_rms_38} represents
a constraint when ${\bar r}$ is fixed to be $3.8\, \mathrm{fm}$\,.
We point out the following two facts in this parameter search.
Firstly, we have negative $E_\nu$ for $V_r< 370\, \mathrm{MeV}$
and complex $\omega_{120}$ for $V_r< 330\, \mathrm{MeV}$, implying that
BEC is unstable for smaller $V_r$ (consequently  smaller $\Omega$). The former is caused by
the negative ``mass" $1/(I-4C_{1102})<0$ in Eq.~(\ref{eq:HuQP}).
The latter is the dynamical instability 
\cite{Pu1999,Garay2000,Skryabin2000,Wu2003,Mettenen2003,Kawaguchi2004,Mine2007}, and occurs, because the weak repulsive 
interaction cannot prevent BEC from collapsing.
Secondly,  $\Omega$ is the most significant parameter to determine the energy level spacing,
and large $\Omega$ ($> 3\,\mathrm{MeV}$) cannot reproduce 
the observed energy levels. After all, no solution is found for the small ${\bar r}$ that
requires large $\Omega$.
We therefore advance our calculations, taking the maximum ${\bar r}=3.8\, \mathrm{fm}$. 
Fig.~\ref{fig:ex_E_fixed_rms_38} indicates the $V_r$-dependence of calculated energy levels.
Choosing the best fitting parameters $\Omega=2.58\,\mathrm{MeV}$ and $V_r=400\,\mathrm{MeV}$,
called the parameter set B, we give the results of calculated energy levels 
in Fig.~\ref{fig:energyB}. The zero energy spacing are narrower due to the 
smaller $V_r$ and the BdG energy
spacing is wider due to the larger $\Omega$ than those in Fig.~\ref{fig:12Cenergy}. 
As a result, the NGH state with $\nu=4$ is located near $4^+_1$, while the energy level
of the NGH state with $\nu=3$ falls down to the midpoint between $0_4^+$ and $4^+_1$,
and the calculated BdG excitation levels tend to be above the observed levels.


Our interpretation of the  $\alpha$ cluster states as phase locking
due to BEC is quite different from the traditional $\alpha$  cluster model, 
{\it ab initio} calculations, and other approaches 
that try to explain them as collective modes in {\it configuration space\/},
e.g., the rotational band or vibrational states 
caused by breakdown of rotational or translational symmetries.

\par
In the traditional models, there has been a long-standing question 
about which excited states are the
rotational band members built on the Hoyle state \cite{Freer2014}. 
In other words,  which of   the $0_2^+$  and 
 $0_3^+$ states  is the bandhead  of the  observed $2_2^+$  
and   $4_1^+$ states?  
The first and   traditional $\alpha$ cluster model picture regards the Hoyle state as the
 bandhead state
 \cite{Freer2012B,Ogloblin2014,Morinaga1956}.
In the $\alpha$ condensate model \cite{Yamada2005}, the  
 $2_2^+$ state is interpreted as a state in which an $\alpha$ cluster is
 lifted from the Hoyle state 
to the $D$ state in {\it configuration space},  and both states have  essentially the 
 same weakly coupled [$^8$Be($0^+)\times\alpha$]$_{J}$  
 cluster configuration revealed in Refs.\cite{Uegaki1977,Kurokawa2007}.
In these cluster model pictures, because the Hoyle   and 
 $2_2^+$ states have a gas-like spherical structure, it is difficult  to consider logically
that a rotational band is built.
 In {\it ab initio} lattice \cite{Epelbaum2012} and no-core shell model
 \cite{Dreyfuss2013} calculations,
 the Hoyle, $2_2^+$, and $4_1^+$ states   are understood to be  rotational band  
states.
The second  interpretation is that the $0_3^+$  state is a bandhead state on which 
the rotational   $2_2^+$  and $4_1^+$ states  are    built  \cite{Freer2012B}. 
Ref.\cite{Kurokawa2007} suggests that the $0_3^+$ state is a higher nodal
 state with the [$^8$Be($0^+)\times\alpha$($L=0)$]$_{J=0}$ structure. 
A  calculated large $B(E2)$ value of the 2$_2^+$$\rightarrow$0$_3^+$ transition 
 \cite{Kanada2007} is reported, although 
 no experimental data are available. In the next section, we will calculate
the transition probability and also obtain a large value in our approach.

\par
The  reason that these two different  interpretations have been presented  is entirely
 due to the appearance of   the     Hoyle  
and  $0_3^+$ states so closely   above the $\alpha$ threshold.
 If  rotational invariance 
  of the  $0_2^+$ and  $0_3^+$ states in configuration space is broken, a rotational band
 should appear individually on both the  $0_2^+$ and  $0_3^+$ states,
 in contradiction with  the experimental data.
It  seems difficult to determine which interpretation is correct as long as these are considered 
as collective  modes with $\alpha$ cluster structure in  {\it configuration space}.
In our picture above, the question does not  arise in principle.
Our calculations show that the $2_2^+$ and $4 _1^+$  states are the BdG 
%excited BEC phonon 
 states
and need not be  rotational member states on either the Hoyle state or the $0_3^+$ state.
In fact, the $J(J+1)$  plot  of the excitation  energy of the observed states of  the 
 band based on  the above two  pictures deviates from a straight line. 

\par
Why and how does nature allow 
in principle  the emergence of the  $0_4^+$  state, which is interpreted as
  a linear chain-like $\alpha$ 
cluster state in Refs.\cite{Kurokawa2007,Chernykh2007,Kanada2007,Suhara2014},
so  close to the  $0_3^+$   
and     $0_2^+$ states?
 In our picture, the close $0_3^+$  and $0_4^+$  states emerge 
 naturally and fundamentally as the NGH states,
which is a logical  consequence of  BEC of   the Hoyle state,
and the three are  closely interrelated.

\section{$\gamma$ decay}\label{sec-gamma}

We can calculate the $\gamma$ decay transitions, using the wave functions 
that have already been obtained. Below the transitions 
 2$_2^+$$\rightarrow$0$_2^+$ and  2$_2^+$$\rightarrow$0$_3^+$ are considered.

The  interaction of $\alpha$ particle, treated as a point-like particle with a charge $2e$,
with the photon field ${\hat {\bm A}}$, 
is introduced from the gauge principle ${\bm \nabla} \rightarrow {\bm \nabla}-2i e {\hat {\bm A}}$
in the Hamiltonian (\ref{Hamiltonian}), and the interaction Hamiltonian is given by
\begin{align}
{\hat H}_A &\simeq
 - \intx {\hat {\bm j}}(x) \cdot {\hat{\bm A}}(x)\label{eq:hatHA} \, ,\\
{\hat {\bm j}}(x) &=\hpsi^\d(x)\frac{2e}{im}{\bm \nabla} \hpsi(x)
\label{eq:hatj}
\,.
\end{align}
We make a multipole expansion of ${\hat {\bm A}}$. The transitions 
2$_2^+$$\rightarrow$0$_2^+$ and  2$_2^+$$\rightarrow$0$_3^+$ are electric quadrupole transitions, 
and the decay rate for a general electric transition with a photon angular momentum $J$ is
\begin{align}
&{\Gamma}_{fi}({\rm E}:k,J,M)=\frac{8\pi(J+1)}{J((2J+1)!!)^2}k^{2J+1}
\notag \\
&\qquad \times
\left| \bra{f}{\hat {\cal M}}({\rm E}:kJM)\ket{i}\right|^2\,,
\label{eq:Gammafi}
\end{align}
where $\ket{i}$, $\ket{f}$ represent the initial and final states of th nucleus with respective
energies, $E_i$ and $E_f$, and $k= E_i-E_f$ is the photon energy. The multipole moment 
${\hat {\cal M}}$ \cite{Bohr1969} is
\begin{align} 
&{\hat {\cal M}}({\rm E}:kJM) =\frac{(2J+1)!!}{k^{J+1}}
\sqrt{\frac{J}{J+1}}\notag \\
&\quad \times \intx {\hat {\bm j}}(\bx)\cdot {\bm \nabla}\times
\left\{ j_J(kr){\bm Y}_{JJM}(\theta,\varphi)\right\} \,,
\end{align}
where $j_\ell$ and ${\bm Y}_{JJM}$ are the spherical Bessel function
and vector spherical harmonics, respectively. When the initial nuclear state is unpolarized
and a sum over the final polarization states is taken, the decay rate is
\begin{align}
&{\bar \Gamma}_{fi}({\rm E}:k,J)
=\frac{8\pi(J+1)}{J((2J+1)!!)^2}k^{2J+1}B({\rm E}J:J_i\, \rightarrow\, J_f) \notag \\
&B({\rm E}J:J_i\, \rightarrow\, J_f) =\frac{1}{2J_i+1} \notag \\
&\qquad \times 
\left|\bra{f(J_f)} \big|
{\hat {\cal M}}({\rm E}:kJ)\big| \ket{i(J_i)}\right|^2 \,,
\end{align}
where $J_i$ and $J_f$ are the initial and final nuclear spins, respectively,
and $B$ is the reduced transition probability \cite{Bohr1969}.

\begin{figure*}[tbh!]
	\includegraphics  [width=17.5cm] {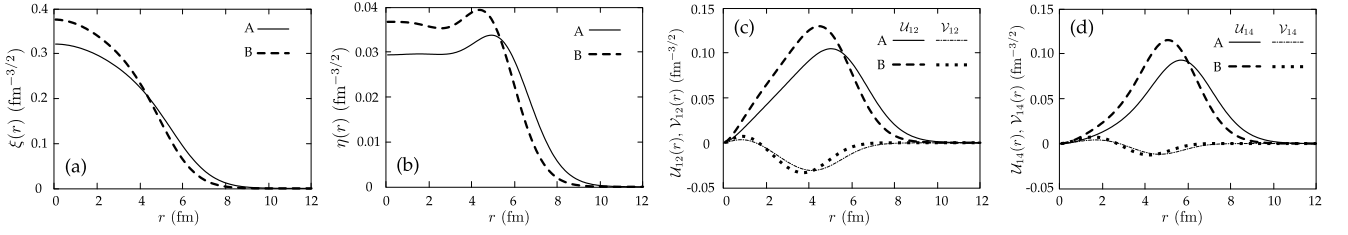}
	\caption{Numerical solutions of (a) $\xi(r)$, (b) $\eta(r)$, (c) $\mathcal{U}_{12}(r)\,,\mathcal{V}_{12}(r)$, 
			and (d) $\mathcal{U}_{14}(r)\,,\mathcal{V}_{14}(r)$ for the parameter sets A and B.
	}
	\label{fig:WaveFunctions1}
\end{figure*}

\begin{figure}[tbh!]
	\includegraphics  [width=8.5cm] {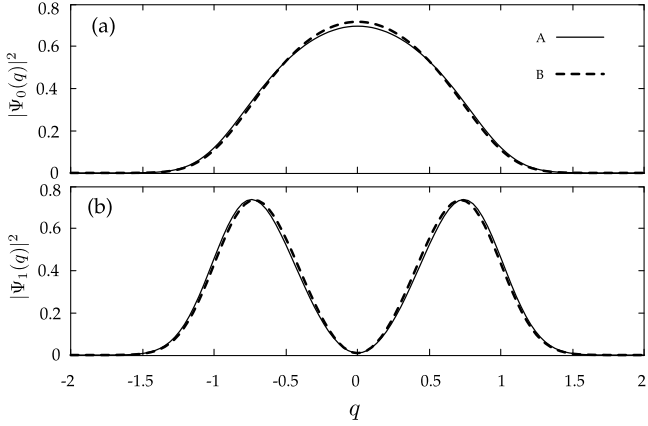}
	\caption{Calculated $|\Psi_\nu(q)|^2$ for the parameter sets A and B.
	}
	\label{fig:WaveFunctions2}
\end{figure}

We calculate ${\bar \Gamma}_{fi}$ and $B$ for the transitions 
2$_2^+$$\rightarrow$0$_2^+$ and  2$_2^+$$\rightarrow$0$_3^+$\,. The states 0$_2^+$,
0$_3^+$, and 2$_2^+$ are identified as the vacuum $\ket{\Psi_0} \ket{0}_\ex$, 
the NGH state $\ket{\Psi_1} \ket{0}_\ex$, and the BdG state 
$\ket{\Psi_0} ({\hat a}^\dagger_{02m} \ket{0}_\ex)$, respectively. Substituting $\hpsi=\xi
+ \hphi$ in Eq.~(\ref{fieldexpansion}) into ${\hat {\bm j}}$ in (\ref{eq:hatj}),
we have the following matrix elements,
\begin{align}
&\bra{f(J_f=0,M_f=0)} 
{\hat {\cal M}}({\rm E}:k 20) \ket{i(J_i=2,M_i=0)} \notag \\
&= \frac{30 e}{im k^3}\sqrt{\frac 23}
\intx \bra{\left\{\begin{matrix}
\Psi_0 \\
\Psi_1
\end{matrix}
\right\}}
\left\{\left(1+i \hat Q\right)  \xi(\bx) + \hat P \eta(\bx)\right\} \notag \\
& \quad \times{\bm \nabla} u_{120} (\bx)
 +v_{120}(\bx){\bm \nabla}\left\{
\left(1-i \hat Q\right)  \xi(\bx) + \hat P \eta(\bx)\right\}
\notag \\
&\quad \ket{\Psi_0} \cdot 
{\bm \nabla}\times\left\{ j_2(kr){\bm Y}_{220}(\theta,\varphi)\right\} \,,
\end{align}
which are further simplified for 2$_2^+$$\rightarrow$0$_2^+$ as
\begin{align}
&\bra{f(0,0)} 
{\hat {\cal M}}({\rm E}:k 20) \ket{N(2,0)} \notag \\
& \qquad = \frac{60e}{mk^3} \int\!dr\, r \left[
\xi (r) \left\{ \frac{d}{dr}  \mathcal{U}_{12}(r)j_2(kr) \right. \right.\notag \\
&\qquad \qquad\quad \left. + \mathcal{U}_{12}(r)
 \left( \frac{j_2(kr)}{r} + kj^\prime_2 (kr)  \right) \right\} \notag \\ 
&\qquad\quad\qquad \left. +   \frac{d}{d r}\xi (r)\mathcal{V}_{12}(r) j_2(kr)   \right]\,,
\end{align}
and for 2$_2^+$$\rightarrow$0$_3^+$ as
\begin{align}
&\bra{f(0,0)} 
{\hat {\cal M}}({\rm E}:k 20) \ket{i(2,0)} \notag \\
&=\frac{60e}{mk^3} \int\!dr\, r \left[
\left\{i  \bra{\Psi_1} \hat{Q}\ket{\Psi_0}  \xi(r) + \bra{\Psi_1}\hat{P}\ket{\Psi_0}
 \eta(r) \right\} \right. \notag \\
&  \times 
\left\{\frac{d}{dr}  \mathcal{U}_{12}(r)j_2(kr)  + \mathcal{U}_{12}(r)
 \left( \frac{j_2(kr)}{r} + kj^\prime_2 (kr)  \right) \right\}
\notag \\
&  + \left\{ i  \bra{\Psi_1} \hat{Q}\ket{\Psi_0} \frac{d}{dr}\xi(r)
 + \bra{\Psi_1}\hat{P}\ket{\Psi_0}
\frac{d}{dr} \eta(r) \right\}  \notag \\
& \qquad\left.  \times \mathcal{V}_{12}(r) j_2(kr)  \right]\,.
\end{align}
Here note that $\bra{\Psi_0} {\hat Q} \ket{\Psi_0}=\bra{\Psi_0} {\hat P} \ket{\Psi_0}=0$
but $\bra{\Psi_1} {\hat Q} \ket{\Psi_0}\,,\,
\bra{\Psi_1} {\hat P} \ket{\Psi_0}\neq 0$, and that the radial functions are defined as
\begin{align}
&\left\{
\begin{matrix}
\xi(\bm{x})\\
\eta(\bm{x})
\end{matrix}
\right\}=\left\{
\begin{matrix}
\xi(r)\\
\eta(r)
\end{matrix}
\right\} Y_{00} (\theta,\varphi ) \,, \notag \\
&\left\{
\begin{matrix}
u_{nJM}(\bm{x})\\
v_{nJM}(\bm{x})
\end{matrix}
\right\}=\left\{
\begin{matrix}
\mathcal{U}_{nJ}(r)\\
\mathcal{V}_{nJ}(r)
\end{matrix} \right\}
 Y_{JM} (\theta,\varphi )  \,.
\end{align}
Using the numerical solutions, $\xi(r)\,,\,\eta(r), \mathcal{U}_{12}(r)\,,\,
 \mathcal{V}_{12}(r)$ and $\Psi_\nu(q)=\braket{q}{\Psi_\nu}$ ($\nu=0,1$)
for each of the parameter sets A and B,
we obtain the reduced transition probabilities that are summarized in Table~\ref{table:B}.
The solutions for each parameter set are shown in Figs.~\ref{fig:WaveFunctions1} and \ref{fig:WaveFunctions2}.

\begin{table}[tbh]
\caption{Calculated reduced transition probabilities $B({\rm E}2:2\, \rightarrow\, 0)$ 
	in unit of $e^2\,{\rm fm}^4$: Ref.~\cite{Kanada2007}, Ref.~\cite{Funaki2015}, and our
	results for the parameter sets A and B.}\label{table:B}
	\begin{ruledtabular}
		\begin{tabular}{ccccc}
			Transition & Ref.~\cite{Kanada2007} & Ref.~\cite{Funaki2015} &Ours (A)& Ours (B)
			\\\hline
			$2_2^+\, \rightarrow \, 0_2^+$ & 100 & 295-340&290& 204 \\
			$2_2^+\, \rightarrow \, 0_3^+$ & 310 & 88-220&342&187\\
		\end{tabular}
	\end{ruledtabular}
\end{table}

It is remarked that the process $2_2^+\, \rightarrow \, 0_3^+$
is the transition between the NGH states, whereas the process $2_2^+\, \rightarrow \, 0_2^+$
is the transition between the BdG states.
The physical picture of condensation in our approach implies that 
the widths of the wave functions, especially 
$\eta(r)$ and $\mathcal{U}_{12}(r)$, are large. But the final results of 
$B({\rm E}2:2\, \rightarrow\, 0)$  are comparable with those in the other
calculations, as in Table~\ref{table:B}.

\section{Summary}\label{sec-Summary}
To summarize,
we have studied the   $\alpha$ cluster structure
 above the $\alpha$ condensate Hoyle state in $^{12}$C by formulating an effective
 field theory  of $\alpha$ cluster condensation that 
properly treats
  spontaneous symmetry breaking of the 
global phase. The observed well-developed $\alpha$ cluster states,
i.e., the $0_3^+$ (9.04 MeV), $2_2^+$ (9.75 MeV), $0_4^+$ (10.56 MeV), and
$4_1^+$ (13.3 MeV) states, are well reproduced. 
Then, the emergence of the NGH states
just above  the Hoyle state is essential. The fact that excitation energies of
the BdG and NGH states are the almost same order of magnitude in our calculation
is also important for the energy spectrum of $^{12}$C.
We adopted the two parameter sets, and both are consistent with the observed
spectrum that has large widths of the energy levels.

We also calculated the $\gamma$ transitions, using the obtained wave functions.
Our results of the reduced transition probabilities are compared with those of the
other model calculations, and are consistent with the latter.

Although the $\alpha$ cluster condensation involves the small number of $\alpha$
particles, it is stable in our study. This is not true in general, 
and actually, when the repulsive 
interaction is weak, we have negative energy
of the NGH state and complex energy of the BdG state that indicate 
an instability of the condensation.

It would be also intriguing to study the  NGH states 
 in other nuclei such as $^{16}$O, $^{20}$Ne, and $^{40}$Ca.

\begin{acknowledgments}
We thank Yasuhiro Nagai and Ryo Yoshioka for their numerical calculations, and
the Yukawa Institute for Theoretical Physics at Kyoto University, 
where our collaboration started at the YITP workshop 
YITP-W-13-13 on ``Thermal Quantum Field Theories and Their Applications.''
This work is partially supported by JSPS KAKENHI Grant Nos. 25400410, 16K05488, 
and by a Waseda University Grant for Special Research Projects (Project No. 2014S-080).
\end{acknowledgments}

%%%%%%%%%%%%%%%%%%%%%%%%%%%%%%%%%%%%%%%%%%%%%%%


\begin{thebibliography}{00}
\bibitem {Cornel2002}
E. A. Cornell and C. E. Wieman, 
Rev. Mod. Phys. {\bf 74}, 875 (2002).

\bibitem{Uegaki1977}
E. Uegaki, S. Okabe, Y. Abe, and H. Tanaka,
 Prog. Theor. Phys. {\bf 57}, 1262 (1977); 
E. Uegaki, Y. Abe, S. Okabe, and H. Tanaka, Prog. Theor. Phys. {\bf 62}, 1621 (1979).

\bibitem {Tohsaki2001}
A. Tohsaki, H. Horiuchi, P. Schuck, and G. Ropke,
 Phys. Rev. Lett. {\bf 87}, 192501  (2001).

\bibitem {Yamada2004} 
T. Yamada and P. Schuck,
 Phys. Rev. C {\bf 69}, 024309  (2004). 

\bibitem {Ohkubo2004}
 S. Ohkubo and  Y. Hirabayashi,
Phys. Rev. C {\bf 70}, 041602 (R) (2004);
Sh. Hamada, Y. Hirabayashi, N. Burtebayev, and S. Ohkubo, 
 Phys. Rev. C {\bf 87},  024311 (2013).

\bibitem {Yamada2005}
  T. Yamada and  P. Schuck, Eur. Phys. J. {\bf A 26}, 185 (2005).

\bibitem {Kurokawa2005} 
C. Kurokawa and K. Kat\=o,
Phys. Rev. C {\bf 71},  021301 (2005);
Nucl. Phys.  {\bf A 738}, 455c (2004). 

\bibitem {Kurokawa2007} % 12C  
C. Kurokawa and K. Kat\=o,
Nucl. Phys.  {\bf A 792}, 87 (2007). 

\bibitem {Kanada2007} % 12C  
Y. Kanada-En'yo,  Prog. Theor.  Phys. {\bf 117}, 655 (2007).

\bibitem {Chernykh2007} 
M. Chernykh, H. Feldmeier, T. Neff, P. vonNeumann-Cosel, and A. Richter,
Phys. Rev. Lett. {\bf 98}, 032501 (2007).

\bibitem {Roth2011}%Structure and Rotations of the Hoyle State Energy 
R. Roth, J. Langhammer, A. Calci, S. Binder, and P. Navratil,
 Phys. Rev. Lett. {\bf 107}, 072501 (2011).

\bibitem {Dreyfuss2013}%sipletic model  Rotations of the Hoyle State Evgeny 
A. C. Dreyfuss, K. D. Launey, T. Dytrych, J. P. Draayer, and C. Bahri, 
Phys. Lett. {\bf B 727}, 515 (2013).

\bibitem {Epelbaum2012}%Structure and Rotations of the Hoyle State Evgeny 
E. Epelbaum, H. Krebs, T. A. Lahde, D. Lee, and Ulf-G. Meissner,
Phys. Rev. Lett. {\bf 109}, 252501 (2012). 

\bibitem {Freer2009} % observation of 2+ state
M. Freer {\it et al.}, 
%, H. Fujita, Z. Buthelezi, J. Carter, R. W. Fearick,
% S. V. F\"{o}rtsch, R. Neveling, S. M. Perez, P. Papka, F. D. Smit,
%J. A. Swartz,and I. Usman,
Phys. Rev. C {\bf 80}, 041303(R) (2009).


\bibitem {Itoh2011} 
M. Itoh {\it et al.}, Nucl. Phys. {\bf A 738}, 268 (2004);
M. Itoh  {\it et al.}, Phys. Rev. C {\bf  84},  054308 (2011).

%12C   second 0+ band states 
\bibitem {Zimmerman2011} 
W. R. Zimmerman, N. E. Destefano, M. Freer, M. Gai, and F. D. Smit,
Phys. Rev. C {\bf 84}, 027304 (2011).

\bibitem {Zimmerman2013}% Unambiguous Identification of the Second 2t State in 12C and the Structure of the Hoyle State
W. R. Zimmerman {\it et al.}, Phys. Rev. Lett. {\bf 110}, 152502 (2013).
\bibitem {Itoh2013} 
M. Itoh {\it et al.},
 J. Phys.: Conf. Ser. {\bf 436}, 012006 (2013).
%\bibitem {Hamada2013}
%Sh. Hamada, Y. Hirabayashi, N. Burtebayev, and 
%S. Ohkubo, Phys. Rev. C {\bf 87}, 024311 (2013).
\bibitem {Freer2011} % observation of 4+ state
M. Freer  {\it et al.}, Phys. Rev. C {\bf 83}, 034314 (2011).
\bibitem {Ogloblin2014} %Rotational band in 12C based on the Hoyle state
A. A. Ogloblin  {\it et al.}, 
EPJ Web Conf.  {\bf 66}, 02074 (2014).
\bibitem {Freer2014} 
M. Freer and H. O. U. Fymbo, Prog. Part. Nucl. Phys. 
{\bf 78}, 1 (2014).


%Makro 4\bibitem {Hamada2013} 
%Sh. Hamada, Y. Hirabayashi, N. Burtebayev, and S. Ohkubo,
% Phys. Rev. {\bf C 87},   024311  (2013).

%\bibitem {Freer1994} 
%M. Freer  {\it et al.}, 
% Phys. Rev. {\bf C 49},  R1751 (1994).
\bibitem {Ring1980}
P.~Ring and P.~Schuck,
{\it The Nuclear Many-body Problem}, (Springer-Verlag, Berlin, 1980).

\bibitem{BalaizotRipka}
J.~P.~Blaizot and G.~Ripka, 
{\it Quantum Theory of Finite Systems},  (MIT press, Cambridge, 1985).

\bibitem {Watanabe2012}
H. Watanabe and H. Murayama,
 Phys. Rev. Lett. {\bf 108},  251602 (2012);
Y. Hidaka, Phys. Rev. Lett. {\bf 110}, 091601 (2013).
\bibitem{Nambu1961}% 
Y. Nambu and G. Jona-Lasinio,
Phys. Rev.  {\bf 122}, 345 (1961).
\bibitem{Goldstone1961}
J.~Goldstone,
Nuovo Cimento {\bf 19}, 154 (1961).
\bibitem {Kadowaki1998}
K. Kadowaki, I. Kakeya, and K. Kindo,
Europhys. Lett. {\bf 42}, 203 (1998).
\bibitem {Littlewood1981}
 P. B. Littlewood and C. M. Varma, Phys. Rev. Lett. {47}, 811 (1981);
P. B. Littlewood and C. M. Varma, Phys. Rev. B {\bf 26}, 4883 (1982).
\bibitem {Varma2001}
 C. M. Varma, J. Low Temp. Phys.  {126}, 901 (2001).
\bibitem{Broglia1973} 
R. A. Broglia, O. Hansen, and C. Riedel, Adv. Nucl. Phys. {\bf 6}, 259 (1973);
R. A. Broglia, J. Terasaki, and  N. Giovanardi,
Phys. Rep. {\bf 335}, 1 (2000);
 D. M. Brink and R. A. Broglia,
{\it Nuclear Superfluidity: Pairing in Finite Systems} 
 (Cambridge University Press, Cambridge, 2005).
\bibitem{ATLAS2012} 
ATLAS Collaboration, Phys. Lett. {\bf B 716}, 1 (2012);
CMS Collaboration, Phys. Lett. {\bf B 716}, 30 (2012).
\bibitem {Matsunaga2013}
R. Matsunaga, Y. I. Hamada, K. Makise, Y. Uzawa, H. Terai, Z. Wang, and R. Shimano,
Phys. Rev. Lett. {\bf 111}, 057002 (2013).
\bibitem {Ohkubo2013} 
S. Ohkubo, arXiv: nucl-th 1301.7485 (2013).
\bibitem{NTY}
Y.~Nakamura, J.~Takahashi, and Y.~Yamanaka, Phys.~Rev.~A {\bf 89}, 013613 (2014).
\bibitem {Ali1966} 
S. Ali and A. R. Bodmer, Nucl. Phys. {\bf A 80}, 99 (1966).
\bibitem{GP}
E.~P.~Gross, Nuovo Cimento {\bf 20}, 454 (1961); 
J.~Math.~Phys. {\bf 4}, 195 (1963);  %\\
L.~P.~Pitaevskii, Zh.~Eksp.~Teor.~Fiz. {\bf 40}, 646 (1961).
\bibitem{Lewenstein}
M.~Lewenstein and L.~You, 
Phys.~Rev.~Lett. {\bf 77}, 3489 (1996).
\bibitem{Matsumoto2}
H.~Matsumoto and S.~Sakamoto, 
Prog.~Theor.~Phys. {\bf 107}, 679 (2002).
\bibitem{Bogoliubov}
N.~N.~Bogoliubov, 
J.~Phys. (Moscow) {\bf 11}, 32 (1947). 
\bibitem{deGennes}
P.~G.~de Gennes, {\it Superconductivity of Metals and Alloys}
(Benjamin, New York, 1966).
\bibitem{Marshlek1969}
E.R.~Marshalek and J.~Weneser, 
Ann.~Phys. {\bf 53}, 569 (1969).
\bibitem {Ogasawara1976}
H. Ogasawara and J. Hiura,  Prog. Theor.  Phys. {\bf 59}, 655 (1978).
\bibitem {Lazauskas2011}
R. Lazauskas and M. Dufour, Phys. Rev. C {\bf 84}, 064318 (2011).
% Momentum distributions of ƒ¿ particles from decaying low-lying C12 resonances
%R. \'Alvarez-Rodriguez, A. S. Jensen, E. Garrido, D. V. Fedorov, and H. O. U. Fynbo,
%Phys.~Rev. {\bf C 77}, 064305 (2008).
\bibitem {Tohsaki1980}
A. Tohsaki-Suzuki, M. Kamimura, and K. Ikeda,  Prog. Theor.  Phys. Suppl.
 {\bf 68}, 359 (1978).
%\bibitem {Danilov2009} %Determination of nuclear radii for unstable states in 12C with diffraction inelastic scattering
%A. N. Danilov {\it et al.}, 
%, T. L. Belyaeva, A. S. Demyanova, S. A. Goncharov,  and A. A Ogloblin,
%Phys. Rev. C {\bf 80},  054603 (2009).

\bibitem{Fukushima1977}
M. Kamimura, Nucl. Phys. {\bf A 351}, 456 (1981).

\bibitem{Matsmura2004}
H. Matsumura and Y. Suzuki, Nucl. Phys. {\bf A 739}, 238 (2004).

\bibitem{Danilov2009}
A. N. Danilov, T. L. Belyaeva, A. S. Demyanova, S. A. Goncharov, and A. A Ogloblin, Phys. Rev. C
{\bf 80}, 054603 (2009).

\bibitem{Pu1999}
H.~Pu, C.~K.~Law, J.~H.~Eberly, and N.~P.~Bigelow, Phys. Rev. A
{\bf 59}, 1533 (1999).

\bibitem{Garay2000}
L.~J.~Garay, J.~R.~Anglin, J.~I.~Cirac, and P.~Zoller,
Phys. Rev. Lett. {\bf 85}, 4643, (2000); Phys. Rev. A {\bf 63}, 023611
(2001).

\bibitem{Skryabin2000}
D.~V.~Skryabin, Phys. Rev. A. {\bf 63}, 013602 (2000).

\bibitem{Wu2003}
B.~Wu and Q.~Niu, New J. Phys. {\bf 5}, 104 (2003). 

\bibitem{Mettenen2003}
M.~M{\" o}tt{\" o}nen, T.~Mizushima, T.~Isoshima, M.~M.~Salomaa, and
K.~Machida, Phys. Rev. A {\bf 68}, 023611 (2003).

\bibitem{Kawaguchi2004}
Y.~Kawaguchi and T.~Ohmi, Phys. Rev. A. {\bf 70}, 043610 (2004).

\bibitem{Mine2007}
 M.~Mine, M.~Okumura, T.~Sunaga and Y.~Yamanaka, Ann. Phys. {\bf 322}, 2327 (2007).

\bibitem {Freer2012B} 
M. Freer, J. Phys.: Conf. Ser. {\bf 436}, 012002 (2013).
\bibitem {Morinaga1956}
H.~Morinaga,
 Phys. Rev.  {\bf 101}, 254 (1956).

\bibitem {Suhara2014}
T. Suhara, Y. Funaki, B. Zhou, H. Horiuchi, and A. Tohsaki,
 Phys. Rev. Lett. {\bf 112}, 062501 (2014).

\bibitem{Bohr1969}
A. Bohr and B. R. Mottelson, {\it Nuclear Structure Vol. I},
(Benjamin, New York, 1969).
\bibitem{Funaki2015}
Y. Funaki, Phys. Rev. C {\bf 92}, 021302(R) (2015).
\end{thebibliography}
\end{document}